\documentclass[a4paper]{jpconf}
\usepackage{iopams}

\newcommand{\halv}{\frac{1}{2}}
\newcommand{\inv}[1]{\frac{1}{#1}}

\newcommand{\pil}{\rightarrow}

\newcommand{\nr}[1]{(\ref{#1})}

\begin{document}

\title{Gravity coupled to a scalar field in extra dimensions}
\author{Ingunn Kathrine Wehus\footnote{This is a more complete version
  of a talk given at the XXIX Spanish Relativity Meeting 
\emph{Einstein's legacy: from the theoretical paradise to
  astrophysical observation} (ERE2006), Palma de Mallorca, Spain,
Sept.\ 4-8,  2006.} 
and Finn Ravndal}
\address{ Department of Physics, University of Oslo,
P.O.Box 1048  Blinderen, 
N-0316 Oslo, Norway.}
\ead{i.k.wehus@fys.uio.no}

\begin{abstract}
In $d+1$ dimensions we solve the equations of motion for the case of
gravity minimally or conformally coupled to a scalar field. For the
minimally coupled system the equations can either be solved directly
or by transforming vacuum solutions, as shown before in $3+1$
dimensions by Buchdahl. In $d+1$ dimensions the solutions have been
previously found directly by Xanthopoulos and Zannias. Here we first
rederive these earlier results, and then extend Buchdahl's method of
transforming vacuum solutions to $d+1$ dimensions. We also review the
conformal coupling case, in which $d+1$ dimensional solutions can be
found by extending Bekenstein's method of conformal transformation of
the minimal coupling solution.  Combining the extended versions of
Buchdahl transformations and Bekenstein transformations we can in
arbitrary dimensions always generate solutions of both the minimal and
the conformal equations from known vacuum solutions.
\end{abstract}

\section{Introduction}
Theories in which gravity couples to a scalar field are common in
extra-dimensional problems. A few examples are Kaluza-Klein, Jordan
and 
Brans-Dicke theories, as well as string
theory in general. Consequently, the corresponding Einstein equations
have been studied extensively for the last sixty years, with
particular emphasis on searching for black hole solutions.

Minimally coupled scalar fields in $3+1$ dimensions were first studied
by Fisher \cite{Fisher} in 1948, and later by Bergmann and Leipnik
\cite{BL} and by Janis, Newman and Winicour \cite{JNW}. These were all
solving the equations directly. However, in 1959 Buchdahl showed that
it is always possible to generate a solution for the minimal coupling
case by means of a particular transformation \cite{Buchdahl} of a
vacuum solution metric. The problem was later revisited by Janis,
Robinson and Winicour \cite{JRW}, who also included electromagnetism
in their solutions.

The extension from $3+1$ dimensions to $d+1$ dimensions was first done
by Xanthopoulos and Zannias \cite{XZ89}, using a direct solution
technique. In the present paper we generalize Buchdahl's
transformation method to arbitrary dimensions to solve the same
problem.

Solutions of the conformally coupled equations in $3+1$ dimensions
were first found by Bocharova, Bronnikov and Melnikov in 1970
\cite{BBM}, using a direct approach. However, these results were
published in a Russian journal, and did therefore not get the
attention of physicists in the western world. As a result, the
solutions were re-discovered independently by Bekenstein
\cite{Bekenstein1} in 1974, using a conformal transformation
method. These solutions included a black hole-like solution
\cite{Bekenstein2}, known as the BBMB
(Bocharova--Bronnikov--Melnikov--Bekenstein) black hole.  The first
direct solution in the West was found by Fr\o yland \cite{Froyland} in
1982, who demonstrated that the metric coincides with the extremal
Reissner-Nordstr\"om solution. In a later work, Xanthopoulos and
Zannias \cite{XZ91} further showed that the BBMB solution is unique.

Again, the extension to $d+1$ dimensions was also done by Xanthopoulos,
this time together with Dialynas \cite{XD}. They found solutions for
the conformally coupled case by extending Bekenstein's method of
transforming minimal solutions. They also demonstrated that the BBMB
black hole only exists in $3+1$ dimensions. Klimcik \cite{Klimcik}
immediately after found these solutions by a direct integration technique.

These issues have been investigated in collaboration with Tangen
\cite{Tangen}. One goal of this investigation was to extend Buchdahl's
and Bekenstein's methods of generating solutions by conformal
transformations to spacetimes with extra dimensions.  However, during
the course of the project, we found that most results had already been
obtained by others.  We will here give a pedagogical introduction to
the field based on the more formal approach of Tangen. In particular,
we will show how the Buchdahl transformation can be extended to
arbitrary dimensions.

\section{Gravity coupled to a scalar field}
\label{sec:grav_scal_field}

In $D=d+1$ dimensions the theory of gravity coupled to a free scalar field is
described by  the action
\begin{equation}\label{virkninga}
S=\int d^Dx\sqrt{-g}\;\halv\Bigl(
R-\xi R\phi^2
-g^{\mu\nu}\phi_{,\mu}\phi_{,\nu} 
\Bigr) 
\end{equation}
We work in natural units where $c=1=M_{D}$, $M_{D}$ being the
$D$-dimensional reduced Planck mass. For instance, in four spacetime
dimensions we have $M_4^{\;-2}={8\pi G_3}=1$.

Putting the parameter $\xi$ to zero gives us the common case where the
scalar field is minimally coupled to gravity, while keeping $\xi$
non-zero allows for more general couplings.  In this paper we will
also be interested in the conformal coupling case defined by
$\xi=\frac{d-1}{4d}$.  Varying the action with respect to $\phi$ gives
the equation of motion for the scalar field
\begin{equation}\label{philign}
\opensquare^2\phi-\xi R\phi=0
\end{equation}
while varying with respect to the metric $g^{\mu\nu}$ gives the
Einstein equations
\begin{equation}\label{einsteinlign}
(1-\xi\phi^2)E_{\mu\nu}=T_{\mu\nu}^{\;\phi}+\Delta T_{\mu\nu}^{\;\phi}
\end{equation}
Here $T_{\mu\nu}^{\;\phi}$ is the ordinary scalar field
energy-momentum tensor for the minimal coupling ($\xi=0$) case
\begin{equation}\label{mintphi}
T_{\mu\nu}^{\;\phi}=
\phi_{,\mu}\phi_{,\nu} 
-\halv g_{\mu\nu }\phi_{,\alpha}\phi^{,\alpha}
\end{equation}
and
$\Delta T_{\mu\nu}^{\;\phi}$ is the Huggins term \cite{Huggins} coming from the
extra term $\xi R\phi^2$ in the Lagrangian \cite{CCJ}
\begin{equation}
\Delta T_{\mu\nu}^{\;\phi}=
\xi\left[g_{\mu\nu}\opensquare^2(\phi^2)-(\phi^2)_{;\mu\nu} \right]
\end{equation}
To summarize, the total energy-momentum tensor is given by
\begin{equation}
T_{\mu\nu}=\phi_{,\mu}\phi_{,\nu} 
-\halv g_{\mu\nu }\phi_{;\alpha}\phi^{;\alpha} 
+\xi\left[g_{\mu\nu}\opensquare^2(\phi^2)-(\phi^2)_{;\mu\nu} \right]
+\xi\phi^2 E_{\mu\nu}
\end{equation}
Only for $\xi=\frac{d-1}{4d}$ will $T_{\mu\nu}$ be traceless and the
theory will be conformally invariant. By contracting equation
\nr{einsteinlign} we see that in this case also the Ricci scalar
vanishes, $R=0$. Taking the trace of equation \nr{mintphi}, we notice
that only in $1+1$ dimensions is a minimally coupled scalar field
conformally invariant.

\subsection{Statical, spherical symmetric solutions}
We are searching for black hole-like solutions and are only interested
in static and spherical symmetrical solutions. The most general static
and spherical symmetric metric in $d+1$ dimensions can be written
\begin{equation}\label{kulemetrikk}
ds^2 = -e^{2\alpha(r)}dt^2 + e^{2\beta(r)}dr^2 
+ e^{2\gamma(r)}r^2d\Omega_{d}^{\;  2}
\end{equation}
where $\alpha$, $\beta$ and $\gamma$ are unknown functions of the
radial coordinate $r$, and $d\Omega_{d}^2$ is the solid angle element
in $d-1$ dimensions.  Putting $\gamma=1$ gives us Schwarzschild
coordinates in which the equations of motion often take the simplest
form. In our case the solutions can often not be explicitly written
down in Schwarzschild coordinates and it is better to work in
isotropic coordinates, defined by $\beta(r)=\gamma(r)$.  In
Schwarzschild coordinates the Einstein tensor has the independent
components
\begin{eqnarray}\label{E00}
e^{2\beta-2\alpha}\,
E_{tt}&=& 
\frac{(d-1)}{r}\beta'
+\frac{(d-2)(d-1)}{2r^2}(e^{2\beta}-1)
\\\label{E11}
E_{rr}&=&\frac{(d-1)}{r}\alpha'
-\frac{(d-2)(d-1)}{2r^2}(e^{2\beta}-1)
\end{eqnarray}
The scalar field configuration must also be static and spherically
symmetric, so $\phi_{,\mu}$ can only have one non-zero component,
$\phi_{,r}\equiv\phi'$.  Using this when calculating the
energy-momentum tensor we find for the minimal case
\begin{equation}\label{Tmin}
e^{2\beta-2\alpha}\,T^{\;\phi}_{tt}=T^{\;\phi}_{rr}=\halv\phi'^2
\end{equation}
and for the Huggins term
\begin{eqnarray}
e^{2\beta-2\alpha}\,
\Delta T^{\;\phi}_{tt}&=& 
-2\xi\left[
\phi\phi''+\phi\phi'\left(\frac{d-1}{r}-\beta'\right)+\phi'^2
\right]
\\
\Delta T^{\;\phi}_{rr}&=& 
2\xi\phi\phi'\left(\frac{d-1}{r}+\alpha'\right)
\end{eqnarray}
We also find the following expression for the D'Alambertian operator
\begin{equation}\label{delphi}
\opensquare^2=
e^{-2\beta}\Bigl[
\partial_r^{\;2}+\bigl(\alpha'-\beta'+\frac{d-1}{r}\bigr)\partial_r  \Bigr]
\end{equation}
Using this we get the following simple expression for the first
component of the Ricci tensor
\begin{equation}\label{R00alpha}
R_{tt}=e^{-2\alpha}\opensquare^2\alpha
\end{equation}
while the Ricci scalar may be written
\begin{equation}\label{ricci}
R=-2\opensquare^2\alpha+\frac{d-1}{r^{d-1}}
\left[r^{d-2}\left(1-e^{-2\beta}\right)\right]'
\end{equation}

\section{Minimal coupling}
\label{sec:mincoup}

\subsection{Fundamental equations}

Putting $\xi=0$ gives us the minimal coupling case. Equation
\nr{philign} then reduces to $\opensquare^2\phi=0$, and using equation
\nr{delphi} we obtain the equation of motion for $\phi$, 
\begin{equation}\label{kulephi}
\phi''=-\bigl(\alpha'-\beta'+\frac{d-1}{r}\bigr)\phi' 
\end{equation}
For a non-constant scalar field this can easily be integrated to give
\begin{equation}\label{kulephilos}
\phi'=C e^{-\alpha+\beta}r^{-(d-1)}
\end{equation}
where $C$ is a constant of integration.  We further notice from
\nr{R00alpha} that $R_{tt}=0$ gives us $\opensquare^2\alpha=0$ so for
both $\alpha$ and $\phi$ non-zero we have
\begin{equation}\label{alphaphi}
\alpha'=K\phi'
\end{equation}
for some constant $K$. When using (\ref{E00}-\ref{E11}) and \nr{Tmin}
the Einstein equations can be simplified to
\begin{eqnarray}\label{enkelein}
e^{2\beta}-1&=&\frac{r}{d-2}\left(\alpha'-\beta'\right)
\\ \label{enkelto}
\phi'^{\;2} &=&\frac{d-1}{r}\left(\alpha'+\beta'\right)
\end{eqnarray}
Using equation \nr{alphaphi} to substitute for $\phi'$, and then
eliminating $\beta$ and $\beta'$ from equations \nr{kulephilos},
\nr{enkelein} and \nr{enkelto} we are left with a first-order
differential equation for $\alpha$.  But this can not be explicitly
solved to find $\alpha$, indicating that the general solution of the
minimally coupled equations can not be explicitly written i
Schwarzschild coordinates. We can however look at two special cases.

First we notice that for constant $\phi$ equation \nr{enkelto} imply
$\alpha'+\beta'=0$. Then equation \nr{enkelein} can be rewritten as
\begin{equation}
\left[r^{d-2}\left(1-e^{-2\beta}\right)\right]'=0
\end{equation}
which may be integrated explicitly, and we end up with the trivial
Schwarzschild vacuum solution for the metric \cite{Schwarzschild} in
$d+1$ dimensions \cite{MP}
\begin{equation}\label{schwarzmet}
ds^2=-\left(1-\frac{B_s}{r^{d-2}}\right)dt^2
+\left(1-\frac{B_s}{r^{d-2}}\right)^{-1}
dr^2+r^2d\Omega_{d}^{\;  2}
\end{equation}
were the integration constant $B_s$ is canonically normalized to give
Newtonian gravity in the large $r$ limit, $B_s=\frac{G_d M}{2(d-2)}$
with $G_d$ being the $d$-dimensional Newtonian gravitational constant,
and M is the mass of the black hole.

Second, for the special case $\alpha=0$, equation \nr{enkelein} can be
rewritten as
\begin{equation}
\left[r^{2(d-2)}\left(1-e^{-2\beta}\right)\right]'=0
\end{equation}
giving
\begin{equation}
e^{-2\beta}=1-\frac{C'}{r^{2(d-2)}}
\end{equation}
Inserting this into \nr{enkelto} we see that for $\phi'^2$ to be
positive, the integration constant $C'$ has to be negative. We put
$C'=-A^2$ and integrate $\phi'$ to get
\begin{equation}
\phi=\pm\sqrt{\frac{d-1}{d-2}}
\ln\left(\sqrt{1+\frac{A^2}{r^{2(d-2)}}}-\frac{A}{r^{d-2}}\right)+C''
\end{equation}
Since we want $\phi$ to go to zero for large $r$ when the metric
approaches flat Minkowski space we choose the integration constant
$C''=0$. Our final solution thus reads
\begin{eqnarray}\label{minlos21}
ds^2&=& -dt^2+\left(1+\frac{A^2}{r^{2(d-2)}}\right)^{-1}dr^2
+r^2d\Omega_{d}^{\;  2}
\\\label{minlos22}
\phi&=&\pm\sqrt{\frac{d-1}{d-2}}
\ln\left(\sqrt{1+\frac{A^2}{r^{2(d-2)}}}-\frac{A}{r^{d-2}}\right)
\end{eqnarray}

Xanthopoulos and Zannias \cite{XZ89} found a general two-parameter
solution in arbitrary dimensions by solving the equations of motion in
isotropic coordinates. This solution may be written as
\begin{eqnarray}
ds^2&=&-\left[\frac{r^{d-2}-r_0^{d-2}}{r^{d-2}+r_0^{d-2}}\right]^{2a} dt^2
+\left[1-\frac{r_0^{2(d-2)}}{r^{2(d-2)}}\right]^{\frac{2}{d-2}}
\left[\frac{r^{d-2}-r_0^{d-2}}{r^{d-2}+r_0^{d-2}}\right]^{-\frac{2a}{d-2}}
\left(dr^2+r^2d\Omega_{d}^{\;  2}\right)
\\
\phi&=&\sqrt{\frac{d-1}{d-2}\left(1-a^2\right)}
\ln\frac{r^{d-2}-r_0^{d-2}}{r^{d-2}+r_0^{d-2}}
\end{eqnarray}
where $r_0$ and $a$ are arbitrary constants.  The parameter $a$ can
run between $0$ and $1$, and $a=1$ corresponds to the Schwarzschild
metric \nr{schwarzmet} plus a constant scalar field solution. $a=0$
gives the upper sign version of (\ref{minlos21}-\ref{minlos22}). As
showed by Xanthopoulos and Zannias this latter solution is a naked
singularity, and no value of $a$ yields a black hole solution
\cite{XZ89}.


\subsection{Buchdahl transformations}
\label{sec:buchdahl}
We will now show that for a given solution of the $d+1$ dimensional
Einstein equations in vacuum, one can always generate a solution of
the same equations minimally coupled to a scalar field.  In $3+1$
dimensions this was first shown by Buchdahl \cite{Buchdahl} and later
by Janis, Robinson and Winicour \cite{JRW}. For the general $d+1$
dimensional case see Tangen \cite{Tangen}.

For a metric on the form 
\begin{equation}\label{Vmetrikk}
ds^2= -e^{2V(x^i)}dt^2+ e^{-2V(x^i)}\hat h_{ij}x^ix^j
\end{equation}
$\hat h_{ij}$ being a $d$-dimensional spatial metric and $V$ is a
function of spatial coordinates only, the Ricci tensor reads
\begin{eqnarray}
R_{00}&=&e^{4V} \left[\hat{\opensquare}^2 V -(d-3)V_{,i}V^{,i}  \right]
=e^{2V}{\opensquare}^2 V
\\
R_{ij}&=&\hat{R}_{ij}
+(d-3)V_{\hat{;}ij}+(d-5)V_{,i}V_{,j}+\hat{h}_{ij}
\left[\hat{\opensquare}^2 V -(d-3)V_{,i}V^{,i}  \right]
\end{eqnarray}
The hat denotes quantities derived using the metric $\hat h_{ij}$
which also is used to raise indexes.  If the metric \nr{Vmetrikk} is a
solution of the Einstein equations in vacuum we must have
$R_{\mu\nu}=0$, which implies
\begin{eqnarray}\label{R00vak}
\hat{\opensquare}^2 V -(d-3)V_{,i}V^{,i}=0
\\ \label{Rijvak}
\hat{R}_{ij}
+(d-3)V_{\hat{;}ij}+(d-5)V_{,i}V_{,j}=0
\end{eqnarray}

We now introduce a new metric
\begin{equation}\label{minmetrikk}
d\bar{s}^2= -e^{2U(x^i)}dt^2+ e^{-2U(x^i)}\tilde h_{ij}x^ix^j
\end{equation}
where again $U$ is a function of spatial coordinates only, and the
spatial metric $\tilde h_{ij}$ is conformal to the spatial vacuum
metric $\hat h_{ij}$. We want our metric \nr{minmetrikk} to be a
solution of the minimally coupled equations. For this metric the
Einstein equations for gravity minimally coupled to a static scalar
field, $\bar{R}_{\mu\nu}=\phi_{,\mu}\phi_{,\nu}$, reads
\begin{eqnarray}\label{R00min}
\tilde{\opensquare}^2 U -(d-3)U_{,i}U^{,i}=0
\\ \label{Rijmin}
\tilde{R}_{ij}
+(d-3)U_{\tilde{;}ij}+(d-5)U_{,i}U_{,j}=\phi_{,i}\phi_{,j}
\end{eqnarray}
while we have the equation of motion for the scalar field $\phi$
\begin{equation}\label{phimin}
\opensquare^2\bar\phi=0=
e^{2U} \left[\tilde{\opensquare}^2 \phi -(d-3)\phi_{,i}U^{,i}  \right]
\end{equation}
Since the Ricci tensor in $d$ dimensions transforms as
\begin{eqnarray}
\hat R_{\mu\nu}
\;\;=\;\;\tilde R_{\mu\nu}
\!\!&-&\!\!
(d-2)\Omega^{-1}\Omega_{\tilde{;}\mu\nu}
-\Omega^{-1}\tilde{h}_{\mu\nu}\tilde{\opensquare}^2 \Omega
\nonumber\\
\!\!&+&\!\!
2(d-2)\Omega^{-2}\Omega_{,\mu}\Omega_{,\nu}
-(d-3)\Omega^{-2}\tilde{h}_{\mu\nu}\Omega^{,\alpha}\Omega_{,\alpha}
\end{eqnarray}
under a Weyl transformation $\hat h_{ij}=\Omega^2\tilde{h}_{ij}$ of
the metric, we find that when setting 
\begin{eqnarray}
U&=&aV
\\
\hat{h}_{ij}&=&\Omega^2\tilde{h}_{ij}
=e^{2\frac{d-3}{d-2}\frac{1-a}{a}U}\tilde{h}_{ij}
\\  \label{phiU}
\phi&=&\pm\sqrt{\frac{d-1}{d-2}\frac{1-a^2}{a^2}}U
\end{eqnarray}
the equations (\ref{R00vak}-\ref{Rijvak}) are transformed into
(\ref{R00min}-\ref{Rijmin}). The equation of motion for $\phi$
\nr{phimin} is also fulfilled.  We see from \nr{phiU} that a necessary
constraint is $a^2\leq1$.  In conclusion, given a vacuum solution of
the Einstein equations on the form \nr{Vmetrikk}, we can always find a
solution of the Einstein equations for a minimally coupled scalar field
given by
\begin{eqnarray}
ds^2&=&-e^{2Va}dt^2+ e^{-2Vb}\tilde h_{ij}x^ix^j
\\
\phi&=&\pm\sqrt{\frac{d-1}{d-2}\left(1-a^2\right)}V
\end{eqnarray}
where $a$ is an arbitrary constant and $b=\frac{a+d-3}{d-2}$.

We adopt the higher-dimensional Schwarzschild solution \nr{schwarzmet}
written in the form
\begin{equation}
ds^2=-e^{2V}dt^2+ e^{-2V}
\left[dr^2+e^{2V}r^2d\Omega_d^{\;2}\right]
\end{equation}
where
\begin{equation}\label{e2V}
e^{V}=\sqrt{1-\frac{B}{r^{d-2}}}
\end{equation}
as our vacuum solution. Using the above transformation, we end up with
the following two-parameter set of solutions
\begin{eqnarray}\label{minmetlos}
ds^2&=&-e^{2Va}dt^2+ e^{-2V\frac{a+d-3}{d-2}}
\left[dr^2+e^{2V}r^2d\Omega_d^{\;2}\right]
\\ \label{minphilos}
\phi&=&\pm\halv\sqrt{\frac{d-1}{d-2}\left(1-a^2\right)}\ln\left(1-\frac{B}{r^{d-2}}\right)
\end{eqnarray}
Renaming the constant $B=4r_0^{d-2}$ and making the coordinate
transformation to isotropic coordinates
\begin{equation}
r\pil r\left(r^{d-2}-r_0^{d-2}\right)^{\frac{2}{d-2}}
\end{equation}
we see when choosing the upper sign in \nr{minphilos} that these are
exactly the same solutions as those found by Xanthopoulos and Zannias
\cite{XZ89}.

\section{Conformal coupling}
\label{sec:concoup}

\subsection{Fundamental equations}
We now consider the case $\xi=\frac{d-1}{4d}$ for which the system is
conformally invariant. Since $T_{\mu\nu}$ , $E_{\mu\nu}$ and
$R_{\mu\nu}$ are traceless in a conformal theory, equation
\nr{philign} still reduces to $\opensquare^2\phi=0$ and the Einstein
equations in Schwarzschild coordinates can be written
\begin{eqnarray}\label{kulein}
\frac{\phi'^{\;2}}{d(d-1)} +\frac{\phi\phi'\alpha'}{d}
&=&
\Bigl(1-\frac{d-1}{4d}\phi^2\Bigr)
\Bigl[
\frac{2}{r}\beta'+\frac{d-2}{r^2}(e^{2\beta}-1)
\Bigr]
\\ \label{kuleinein}
\frac{\phi'^{\;2}}{(d-1)} 
+\frac{\phi\phi'}{d}\left(\alpha'+\frac{d-1}{r}\right)
&=&
\Bigl(1-\frac{d-1}{4d}\phi^2\Bigr)
\Bigl[
\frac{2}{r}\alpha'-\frac{d-2}{r^2}(e^{2\beta}-1)
\Bigr]
\end{eqnarray}
Further, since $\opensquare^2\phi=0$, equations \nr{kulephi} and
\nr{kulephilos} are still valid.

When trying to solve the above equations, it is tempting to choose
$\alpha'+\beta'=0$. Then we get from adding \nr{kulein} and
\nr{kuleinein}
\begin{equation}\label{0011}
\phi\phi''=\frac{d+1}{d-1}(\phi')^2
\end{equation}
with solution
\begin{equation}\label{0011løsning}
\phi=\frac{A}{(r-B)^{\frac{d-1}{2}}}
\end{equation}
where $A$ and $B$ are constants. Combining with 
the $\phi$-equation \nr{kulephilos} which for the case
$\alpha'+\beta'=0$ reads 
\begin{equation}\label{jada}
\phi'=Ce^{-2\alpha}r^{-(d-1)}
\end{equation}
we can solve for $e^{2\alpha}$ and find
\begin{equation}
\label{alphabetaløys}
e^{2\alpha}=e^{-2\beta}
=\Bigl(1-\frac{B}{r}\Bigr)^{\frac{d+1}{2}}r^{-\frac{d-3}{2}}
\end{equation}
Here we have chosen the integration constant $C$ in \nr{jada} such
that the metric approaches Minkowski for large $r$.  But only for
$d=3$ dimensions is this a solution of the Einstein equations
\nr{kulein} and \nr{kuleinein}. In $3+1$ dimensions we find
$A^2=\frac{B^2}{\xi}=6B^2$ and the metric given by \nr{alphabetaløys}
reduces to the $d=3$ version of the extremal Reissner-Nordstr\"{o}m
metric\cite{Reissner,Nordstrom}\cite{MP}
\begin{equation}\label{nordstrom}
ds^2=-\left(1-\frac{B}{r^{d-2}}\right)^2 dt^2
+\left(1-\frac{B}{r^{d-2}}\right)^{-2}dr^2
+r^2d\Omega_d^{\;2}
\end{equation}
The corresponding scalar field solution is
\begin{equation}\label{bbmbphi}
\phi=\sqrt{6}\frac{B}{r-B}
\end{equation}
To get canonical normalization we must put $B=B_s/2$.  For $d\not =3$
the $d+1$-dimensional extremal Reissner-Nordstr\"{o}m metric is not a
solution of the combined gravity and scalar field equations.  In this
case there are no solutions having $\alpha'+\beta'=0$.


\subsection{Bekenstein transformations}

Introducing a new
metric $\tilde{g}_{\mu\nu}$ in the form of the following conformal
transformation of the old metric  ${g}_{\mu\nu}$
\begin{equation}\label{konftrans}
\tilde{g}_{\mu\nu}=
\cosh^{\frac{4}{d-1}}(\sqrt{\xi}\phi)\;
g_{\mu\nu}
\end{equation}
together with a redefinition of the scalar field
\begin{equation}\label{phitrans}
\psi=\inv{\sqrt{\xi}}\tanh(\sqrt{\xi}\phi)
\end{equation}
brings us from the minimal coupling case of the Einstein plus scalar
field equations in the old metric ${g}_{\mu\nu}$
\begin{eqnarray}
E_{\mu\nu}(g)&=&T_{\mu\nu}^{\;\phi}(g) \\
\opensquare^2\phi&=&0
\end{eqnarray}
to the conformal coupling case of the same equations for the new
metric $\tilde{g}_{\mu\nu}$
\begin{eqnarray}
(1-\xi\psi^2)E_{\mu\nu}(\tilde g)&=&T_{\mu\nu}^{\;\psi}(\tilde g) 
+\Delta T_{\mu\nu}^{\;\psi}(\tilde g) \\
\tilde{\opensquare}^2\psi&=&0\;\;=\;\;R
\end{eqnarray}
Here we use the fact that the minimal scalar field energy-momentum
tensor \nr{mintphi} does not change under a conformal transformation
like \nr{konftrans}
\begin{equation}
{T}_{\mu\nu}^{\;\phi}(\tilde g)
={T}_{\mu\nu}^{\;\phi}(g)
\end{equation}
and that the Einstein tensor under a conformal transformation $\tilde
g_{\mu\nu}=\Omega^2 g_{\mu\nu}$ in $D$ dimensions change like
\begin{eqnarray}
\tilde E_{\mu\nu}
\;\;=\;\;E_{\mu\nu}
\!\!&+&\!\!(D-2)\Omega^{-1}
\Bigl[g_{\mu\nu} \opensquare^2\Omega-\Omega_{;\mu\nu}
\Bigr]
\nonumber\\\label{einsteintransf}
\!\!&+&\!\!2(D-2)\Omega^{-2}\Omega_{,\mu}\Omega_{,\nu}
+\halv(D-2)(D-5)\Omega^{-2} g_{\mu\nu}\Omega^{,\alpha}\Omega_{,\alpha}
\end{eqnarray}

The result is the same if we instead of transformations 
(\ref{konftrans}-\ref{phitrans}) use
\begin{eqnarray}
\label{konftrans2}
\tilde{g}_{\mu\nu}&=&
\sinh^{\frac{4}{d-1}}(\sqrt{\xi}\phi)\;
g_{\mu\nu}
\\
\label{phitrans2}
\psi&=&\inv{\sqrt{\xi}}\coth(\sqrt{\xi}\phi)
\end{eqnarray}
The latter correspond to $1-\xi\psi^2$ being negative while the first
transformations (\ref{konftrans}-\ref{phitrans}) is used for positive
$1-\xi\psi^2$.  The transformations (\ref{konftrans}-\ref{phitrans})
and (\ref{konftrans2}-\ref{phitrans2}) were first found in $3+1$
dimensions by Bekenstein \cite{Bekenstein1}. Maeda \cite{Maeda} showed
very generally that Lagrangians with arbitrary couplings between
$\phi$ and $R$ in arbitrary dimensions can always be transformed to a
minimally coupled Einstein Frame theory by means of a conformal
transformation. The specific extension of equations
(\ref{konftrans}-\ref{phitrans}) and
(\ref{konftrans2}-\ref{phitrans2}) to arbitrary dimensions was done by
Xanthopoulos and Dialynas \cite{XD}.

We now take as our minimal solution the solution
(\ref{e2V}-\ref{minphilos}) found in section \ref{sec:buchdahl} which
we write like
\begin{eqnarray}
ds^2&=&-e^{2Va}dt^2+e^{-2Vb} dr^2+e^{2V(1-b)}r^2d\Omega_d^{\;2}\\
\phi&=&\pm\sqrt{\frac{d-1}{d-2}(1-a^2)}V
\end{eqnarray}
Performing the transformations (\ref{konftrans}-\ref{phitrans}) 
we arrive at a two-parameter solution of the conformal equations  
\begin{eqnarray}\label{metkonf}
ds^2&=&\left(\frac{e^{2Vc}+e^{-2Vc}}{2}\right)^{\frac{4}{d-1}}\left[
-e^{2Va}dt^2+e^{-2Vb} dr^2+e^{2V(1-b)}r^2d\Omega^2 \right]
\\ \label{phikonf}
\sqrt{\xi}\psi&=&\tanh (\pm2Vc)=\pm\;\frac{e^{4Vc}-1}{e^{4Vc}+1}
\end{eqnarray}
where the constant $c$ is given by
\begin{equation}
c=\frac{d-1}{4}\sqrt{\frac{1-a^2}{d(d-2)}}
\end{equation}
The conformal solution (\ref{metkonf}-\ref{phikonf}) has the same two
parameters $a$ and $B$ as the minimal solution
(\ref{e2V}-\ref{minphilos}). $V$ is still given by \nr{e2V} and we
still have $b=\frac{a+d-3}{d-2}$.

Choosing now the particular solution given by $c=\inv{4}$,
corresponding to $a=\inv{d-1}$ and $b=\frac{d-2}{d-1}$, the metric
\nr{metkonf} simplifies to
\begin{equation}\label{metrikk_c=1/4}
ds^2=\left(\frac{e^{V}+1}{2}\right)^{\frac{4}{d-1}}\left[
-dt^2+e^{-2V} dr^2+r^2d\Omega^2 \right]
\end{equation}
In order to write this particular solution in Schwarzschild
coordinates, we introduce a new radial coordinate $R$ given by
\begin{equation}\label{Rr}
R=\left(\frac{1+\sqrt{1-\frac{B}{r^{d-2}}}}{2}\right)^{\frac{2}{d-1}}r
\end{equation}
Then the metric \nr{metrikk_c=1/4} can be written as
\begin{equation}\label{metikkR(r)}
ds^2=-\left(\frac{R}{r(R)}\right)^2 dt^2
+\left(
{\frac{2}{d-1}\left(\frac{R}{r(R)}\right)^{\frac{d-1}{2}}+\frac{d-3}{d-1}}
\right)^{-2} dR^2+R^2d\Omega^2 
\end{equation}
where $r(R)$ is given implicitly from \nr{Rr}.  When choosing the
lower sign in \nr{phikonf} the scalar field $\phi$ now takes the form
\begin{equation}\label{phi_1/4}
\sqrt{\xi}\psi=\left(\frac{r(R)}{R}\right)^{\frac{d-1}{2}}-1
\end{equation}
Only in $d=3$ dimensions can \nr{Rr} be solved explicitly to give
\begin{equation}\label{rR3}
\inv{r}=\inv{R}\left(1-\frac{B}{4R}\right)
\end{equation}
so that we can write \nr{metikkR(r)} and \nr{phi_1/4} in Schwarzchild
coordinates
\begin{eqnarray}
ds^2&=&-\left(1-\frac{B}{4R}\right)^2 dt^2
+\left(1-\frac{B}{4R}\right)^{-2} dR^2+R^2d\Omega^2 \\
\psi&=&\sqrt{6}\frac{B/4}{R-B/4}
\end{eqnarray}
which is the same BBMB solution with the extremal Reissner-Nordstr\"om
metric as we found in last section (\ref{nordstrom}-\ref{bbmbphi}).
As demonstrated by Xanthopoulos and Zannias \cite{XZ89} and
Xanthopoulos and Dialynas \cite{XD}, this is the only known black hole
solution for gravity coupled to scalar fields except for the trivial
solutions where $\phi$ is constant and the metric is a vacuum black
hole. The BBMB solution has been extensively studied by for instance
Zannias \cite{Zannias} who has shown it not to have a continuous new
parameter, and therefore not to contradict the {\it no scalar
hair}-theorem \cite{SZ}\cite{AB}\cite{BFM}\cite{Winstanley}.  For an overview
see \cite{Bekenstein3}.

\section{Conclusions}
Given a vacuum solution of the Einstein equations, solutions of the
equations for gravity coupled either minimally or conformally to a
massless scalar field can be generated in arbitrary spacetime
dimensions. To obtain a minimal solution we perform a generalized
Buchdahl transformation on our vacuum metric. To obtain a conformal
solution we perform a generalized Bekenstein transformation on this
minimal solution.

In the search for static and spherical symmetric black hole-like
solutions we choose the Schwarzschild black hole as our seeding
metric. This gives us both Xanthopoulos and Zannias' minimal solutions
and Xanthopoulos and Dialynas' conformal solutions.  It is known that
only in $3+1$ dimensions do these solutions include a black hole,
namely the BBMB black hole where the metric is the extremal
Reissner-Nordstr\"{o}m metric. This makes us wonder what is special
with the four-dimensional spacetime we normally call home.

\ack 
We wish to thank Kjell Tangen and Hans Kristian Eriksen for useful
discussions.  This work has been supported by grant NFR 151574/V30
from the Research Council of Norway.

\end{document}